\documentclass[10pt,conference]{IEEEtran}
\pdfoutput=1
\IEEEoverridecommandlockouts
\usepackage{cite}
\usepackage{amsmath,amssymb,amsfonts}
\usepackage{algorithmic}
\usepackage{graphicx}
\usepackage{textcomp}
\usepackage{xcolor}
\def\BibTeX{{\rm B\kern-.05em{\sc i\kern-.025em b}\kern-.08em
    T\kern-.1667em\lower.7ex\hbox{E}\kern-.125emX}}

\usepackage{url}
\usepackage{multirow}
\usepackage[disable]{todonotes}
\usepackage{caption}
\usepackage[utf8]{inputenc}
\usepackage[frozencache,cachedir=minted-cache]{minted}
\usepackage{svg}
\usepackage{tikz}

\begin{document}

\title{Automated Alert Classification and Triage (AACT): An Intelligent System for the Prioritisation of Cybersecurity Alerts}

\author{\IEEEauthorblockN{Melissa Turcotte}
\IEEEauthorblockA{Sophos\\
melissa.turcotte@sophos.com}
\and
\IEEEauthorblockN{Fran\c cois Labr\`eche}
\IEEEauthorblockA{Sophos\\
francois.labreche@sophos.com}
\and
\IEEEauthorblockN{Serge-Olivier Paquette}
\IEEEauthorblockA{Flare\\
}}


\maketitle

\begin{abstract}
Enterprise networks are growing ever larger with a rapidly expanding attack surface, increasing the volume of security alerts generated from security controls. Security Operations Centre (SOC) analysts triage these alerts to identify malicious activity, but they struggle with alert fatigue due to the overwhelming number of benign alerts. Organisations are turning to managed SOC providers, where the problem is amplified by context switching and limited visibility into business processes.

A novel system, named AACT, is introduced that automates SOC workflows by learning from analysts' triage actions on cybersecurity alerts. It accurately predicts triage decisions in real time, allowing benign alerts to be closed automatically and critical ones prioritised. This reduces the SOC queue allowing analysts to focus on the most severe, relevant or ambiguous threats. The system has been trained and evaluated on both real SOC data and an open dataset, obtaining high performance in identifying malicious alerts from benign alerts.

Additionally, the system has demonstrated high accuracy in a real SOC environment, reducing alerts shown to analysts by 61\% over six months, with a low false negative rate of 1.36\% over millions of alerts.
\end{abstract}


\newcommand{\Secureworks}{Sophos}
\newcommand{\Taegis}{Taegis}
\newcommand{\vectorentity}{x}
\newcommand{\totalalertsprefiltering}{$1,310,693$}
\newcommand{\totalalertstesttenantsremoved}{$1,202,868$}
\newcommand{\totalinvestigatedalerts}{$161,886$}
\newcommand{\totalinvestigations}{$30,194$}
\newcommand{\totaltenants}{$1,062$}
\newcommand{\trainingsamplesprioresolvedremoved}{$474,034$}
\newcommand{\trainingsamplespositiveclass}{$161,886$}

\section{Introduction}

A Security Operations Center (SOC) is a team of security professionals, referred to as analysts, that monitor an organisation's IT infrastructure, ranging from cloud applications to networked devices \cite{strategiessoc}. They detect and respond to cybersecurity threats in real time. In place of internally managed SOCs, organisations are increasingly moving towards managed SOCs. The Managed Detection and Response (MDR) market was valued at \$4.9 billion in 2021 with an estimated compound annual growth rate of 18.1\% \cite{mssp_report}. MDR providers use high-availability SOCs to offer 24/7 detection, monitoring, and response for security systems from multiple organisations. For MDR providers and external SOCs, maximising the efficiency of analysts is crucial for scaling across clients while maintaining their security, and reducing the mean time to respond (MTTR) to threats.

SOC analysts have to deal with numerous security alerts from a wide range of security controls, many of which are irrelevant or benign \cite{alahmadi202299}. As alert volumes increase, it becomes harder for analysts to quickly identify severe threats, leading to decreased efficiency as well as delays in incident response, a problem known as alert fatigue.

Managed SOC analysts encounter additional challenges. Each organisation has their own unique network environments, security controls and best practices that are not always exposed to the analyst. Therefore, analysts can lack the business and environmental context for triaging alerts and have to context switch between these different environments daily. This workload can be even larger due to customer support tasks.

To address alert fatigue, the alert life-cycle within a SOC workflow includes multiple stages: collection, reduction, correlation, and prioritisation \cite{wang2023combating, salah13, valeur04}. Collection gathers alerts from various detection systems and normalises them to a standardized format. Reduction filters redundant alerts and does the final validation. Correlation groups alerts that are part of the same attack scenario into security incidents to provide analysts with the necessary context to triage the security threat. Prioritisation ranks alerts by severity before handing them to analysts for review and triage. A final action is taken as the outcome of the triage process, such as escalating the alert for further investigation or deeming the alert benign.

All of these stages are critical in most modern SOC operations, and despite automated techniques \cite{wang2023combating, deepcase, han20unicorn, valeur04}, analysts are still overwhelmed with alerts. 
Shen \textit{et al.}~\cite{shen2018tiresias} make use of a real-world dataset comprising security alerts, with an average rate of 176 security events per device daily, demonstrating the significant number of alerts that companies face each day.

This paper introduces a novel supervised machine learning (ML) system named Automated Alert Classification and Triage (AACT) that prioritises alerts and can automate triage decisions from human analysts. AACT enhances the prioritisation and triage stage of the alert life-cycle by addressing the overload of alerts that analysts face despite upstream efforts to mitigate the issue. The key conjecture is that triage actions on alerts depend on past actions for similar alerts. Automatically learning from how human experts are triaging classes of alerts dynamically over time and using this to automate these decisions was identified as a research gap in the literature.  

To model the human alert triage process, dynamic features are used that encode short-term and long-term trends of triage actions taken over classes of alerts. The short-term trends capture the temporal nature of triage actions that classify alerts as malicious or benign depending on what other threats are immediately present in the network environment. The long-term trends explain the organisational context. For example, long-term noise in some environments, e.g., from scanners, can mean that the presence of a class of alert is often benign, whereas the presence of the same alert in quieter environments is likely to be malicious.

The dynamic features are advantageous as they encode the non-heterogeneities in the data and do not use details of the alert whose relationship with the analyst triage action changes often. Therefore, frequent model retraining is not required. For example, an alert originating from a certain source IP might be malicious over a short time frame, but other times be benign. Instead of explicitly encoding the IP as a feature, the dynamic features used encode how analysts have triaged other alerts with that IP in the recent past.  This results in an approach that can quickly react to trends and is robust to the fast-changing threat landscape. 

These dynamic features, combined with rarity features and static features, are used in a supervised ML model to predict analyst actions for new alerts. The system's predictions can automatically perform triage, prioritise or rank alerts, and suggest actions. The features are understandable by analysts, showing how alert scores relate to how other similar alerts have been triaged.

This flexible framework integrates with any SOC workflow as the model can be easily customized to work with any finite set of analyst triage actions.  Further, AACT has low deployment and update overhead as the features and training data are generated based on the decisions analysts are making as part of the core operation loop of the SOC. 

Finally, the model can adapt to a managed SOC setting, calculating features per customer environment and across all environments. This ensures the model is tailored to each organisation’s unique environment while also learning from global trends enhancing performance.

The framework is evaluated using a 6-month period of real-world data from a managed SOC, as well as a synthetic public dataset. The evaluation includes comparisons against a previous approach from the literature, DeepCase \cite{deepcase}, and a baseline model. In addition, AACT has been deployed in a managed SOC, where over the course of a 6-month period it has processed and helped triage approximately 3.1 million alerts. The system automated the closure of 61\% of alerts while prioritising the remaining ones.

In summary, this paper's contributions are:
\begin{enumerate}
\item Presenting a novel ML framework, AACT, that automates analyst triage decisions by dynamically learning in real-time from their actions using directly interpretable features.
\item Developing an adaptable and easy-to-deploy alert triaging framework that avoids frequent retraining, functions within any standard operating SOC environment without additional overhead, and adapts to various workflows for both managed and in-house SOCs.
\item Demonstrating the framework's effectiveness in a real managed SOC, significantly reducing alert fatigue.
\end{enumerate}

The paper is organised as follows: Section~\ref{sec:relevantwork} discusses related literature; Section~\ref{sec:alertdata} describes the alert data set used to demonstrate the methodology; Section~\ref{sec:methodology} presents the approach to model and predict triage actions; Section~\ref{sec:results} evaluates the approach on a closed and an open dataset, with a comparison to Deepcase; Section~\ref{sec:live-implementation} discusses the live environment performance, and Section~\ref{sec:discussion} and \ref{sec:conclusion} covers future work and conclusions.

\section{Related work}\label{sec:relevantwork}

Related work has explored reducing alert fatigue by targeting various stages of the SOC alert life-cycle \cite{salah13}. \cite{valeur04} offers a comprehensive framework based on rules and heuristics that addresses all the stages of the life-cycle, evaluating each stage individually. Alert reduction and correlation have been extensively researched in the literature and both rule-based \cite{cuppens02, husak19} and data-driven \cite{chyssler04, deepcase, han20unicorn,nadeem22, haas18, sharif2024drsec} approaches have been proposed.

This paper focuses on the final stages of the alert processing pipeline—prioritisation and triage. Some techniques add necessary contextual data for triage before alerts reach human analysts. Triage automatons using finite state machines were developed in \cite{zhong16} and \cite{zhong19} by collecting the human analysts' investigative processes, such as querying and correlating additional event and alert data.
For automating triage, there is a body of work that aims to classify alerts by learning which events or collection of events are more likely to lead to real security incidents. Methods like NoDoze~\cite{hassan2019nodoze}, UNICORN~\cite{han20unicorn}, and OmegaLog~\cite{hassan20omegalog} use provenance graphs to model alert context with system-level or application-layer events as supplementary data. Anomaly scores assigned to event sequences based on rarity of edges within the graph are used to prioritise alerts. Rapsheet~\cite{hassan2020} extends provenance graphs to EDR-generated alerts without the need for system-level events. Threat scores are generated based on MITRE techniques associated with the alerts. Isolation forests in \cite{aminato19} and \cite{siddiqui19} highlight anomalous alerts, which are considered higher severity. FASTT~\cite{mcelwee17} combines context retrieval and alert classification using a neural network, Elasticsearch, and Kibana. These methods do not learn from how analysts have triaged alerts in the past, and any labels are only used to assess the performance of their models. Note that these methods can often be complimentary to AACT, useful at earlier stages of the alert lifecycle.

Context2Vector~\cite{liu22} and DeepCase~\cite{deepcase} use deep learning to correlate alerts and then classify these groups with supervised learning. DeepCase~\cite{deepcase} correlates alerts within sequences using a deep learning model along with an attention mechanism. These correlated alerts are grouped into similar clusters. A semi-supervised approach is used whereby analysts manually label the clusters and then similar sequences to those within the cluster can be automatically labelled. Context2Vector generates topic representations of alert sequences for human labeling. Both methods require external labeling and do not dynamically integrate historical and current analyst triage actions. DrSec~\cite{sharif2024drsec} also employs deep learning, through the pre-training of a language model (LM) to encode event data of processes. It uses self-supervision and has the advantage of requiring few labeled data points to to identify malicious alerts. 

Similarly to the temporal nature of the features proposed in the paper, \cite{gelman23} develops a ML framework using temporal features to predict analyst actions based on network context. Alert attributes, such as the count of the number of other alerts, and the average severity of those alerts for the alert entity are used as features. They use analyst actions as labels to train their model. Our approach differs in that it encodes analyst actions as features, learning dynamically in real-time from the way analysts triage alerts, providing a model that is reactive to the current activity in the SOC. AlertPro~\cite{wang2023combating} along with context and basic alert attribute features employs history features, binary encodings of past analyst decisions, to re-rank alerts using an active learning algorithm. However, it does not account for the frequency or historical context of these alerts in attacks. \todo{Features unrealistic in a live streaming system}

In comparison, AACT uniquely learns in real time from analysts triaging alerts within a live system without additional labeling. By using analyst actions to label and triage similar alerts, AACT captures the dynamic nature of threats and amplifies analyst decision-making. For example, we don’t assume that sequences of alerts are either always malicious or benign but can change over time with respect to other threats in the network environment. Unlike other approaches, AACT integrates both current and historical triage actions into the alert prioritisation process with near-zero latency, addressing a gap in the literature.

\section{Alert Data}\label{sec:alertdata}
\Secureworks{} offers a managed detection and response (MDR) solution that combines an open cloud-native security platform with security expertise through its SOC analysts. The data used for analysis are the security alerts processed by the security platform that are triaged by an analyst.

A cybersecurity alert is a notification that an organisation's information systems may be compromised or undergoing a cyber attack. It is generated by security tools using collected computer system telemetry, and typically refers to a collection of log events or telemetry that have been deemed malicious and are related in some way to the attack. 

At \Secureworks{}, alerts originate both from internal sensors and detectors operating on enterprise network telemetry as well as those ingested by third party vendors. The alerts are normalised into an alert object. This object contains information about the alert, such as the relevant log events, attack techniques used, creation timestamp, origin, detector and rule that generated the alert, and a hard-coded severity. It reports on entities that are related to the attack as extracted from the original log events. This could be the entity where the attack originates, or the entity impacted by the attack. An entity refers to a unique identifier that may represent a human, machine, program, or other digital object that operates within or interacts with a computing environment. It also contains the customer environment where the alert originated, which will be referred to as the {\it tenant} for the remainder of the paper. An example alert object for a subset of fields is shown in Listing~\ref{fig:alert_example}.
\begin{listing}
\begin{minted}[frame=single,framesep=3mm, fontsize=\footnotesize]{js}
{
    "id":"unique_alert_uri",
    "metadata":{
        "creator":{
            "detector":{..}
            "rule":{..}
        },
        "severity":0.75,
        "title":"Unfamiliar sign-in properties",
        "created_at":{..}
    },
    "entities": [..],
    "tenant_id":"11772",
    "status":"OPEN",
    "investigation_ids":[..],
    "event_ids":[..]
}
\end{minted}
\caption{An example alert object at \Secureworks{}}
\label{fig:alert_example}
\end{listing}

As part of the triage process, analysts review a prioritised list of alerts. Alerts that are considered a real threat and worthy of further review and remediation are placed into an {\it investigation}. An investigation is an additional abstraction over a collection of telemetry, alerts, assets, related searches, alerts and log events that are representative of a single related security event. These investigations are sent to the tenant for review with suggested remediation actions. 
Finally, once alerts have been triaged, they are assigned a label indicating if true malicious activity occurred or if it was benign. Alerts that are not placed into an investigation are immediately labelled as benign. Alerts that are placed into an investigation are generally labelled after review and incident remediation. 
 
The history of labels applied to an alert and the investigation the alert was placed in, if any, are maintained within the alert object. \Secureworks{} analysts are primarily analysing alerts to assess whether they are of significant enough security value to be placed in an investigation. Therefore, the primary goal for the analysis in this paper is to be able to predict whether an alert would be placed in an investigation. If there is high enough confidence that the alert would be closed without being placed into an investigation, the system can perform that action prior to it being added to the analyst queue reducing the number of alerts shown to analysts. 

\begin{figure}[h]
    \centering
    \includegraphics[width=1.0\columnwidth]{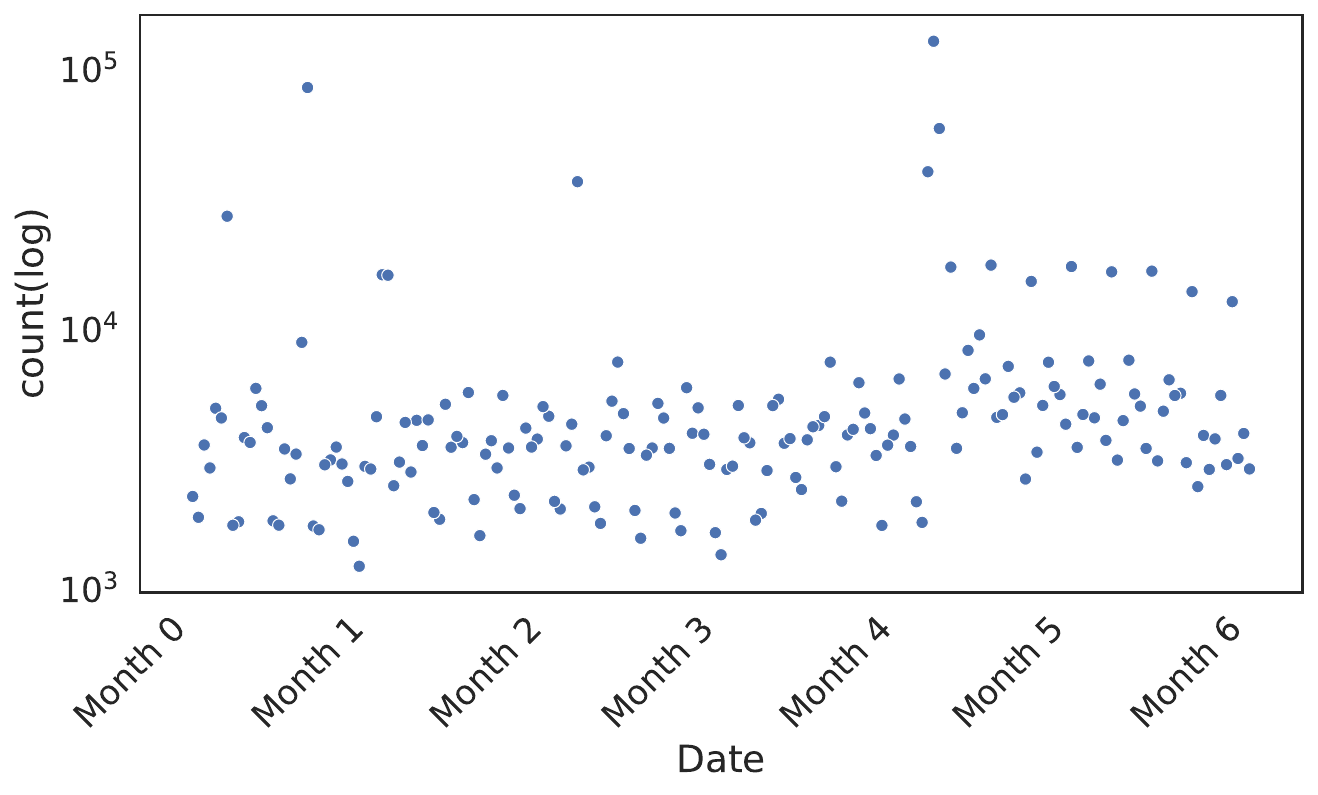}
    \caption{Daily alert counts across all tenants}
\label{fig:alert_spikes}
\end{figure}

The specific dataset used to train and evaluate the methodology presented over the next sections are the alerts and investigations described above over a 6-month period. Note that these alerts have already undergone processing steps, namely de-duplication and rule-based filtering, as part of the alerting lifecycle at \Secureworks{}. Any alerts that arose from test or demo tenants were filtered to remove artificial unrealistic data that can impact building statistical models. There are \totalalertstesttenantsremoved{} alerts and \totalinvestigatedalerts{} alerts were investigated across \totalinvestigations{} unique investigations for multiple tenants in the original dataset. Figure \ref{fig:alert_spikes} shows the daily alert counts across all tenants. As can be seen there are some days where large alert spikes can occur; these are often caused by errors in the alerting pipeline due to missed de-duplication efforts or misconfigured rules or detectors. 

\subsection{Alert similarity}\label{sec:alertsimilarity}

The model used requires a notion of alert similarity so that any alert can be compared to actions taken over other similar alerts that came previously. Two alerts are considered similar if the alert notification is for the same underlying attack behaviour {\it or} there is at least one overlapping entity within the alert belonging to the same tenant. To provide intuition, if an alert with a specific user account was marked by an analyst as malicious, indicating that the user account has been compromised, then future alerts within a similar time frame arising from that user account are probably worthy of investigation. Similarly, if there is an alert type that has been frequently marked as malicious, then it is also likely that future alerts of that type could also be malicious and should have higher priority.

\subsubsection{Alert categories}\label{sec:alertcategories}
To quantify if the same attack behaviour is being detected, alert categorisation is performed. It is assumed that each alert belongs to a specific discrete category related to the signature or attack behaviour being detected, similar to MITRE tactics and techniques\footnote{https://attack.mitre.org/matrices/enterprise/}. As the alerts both originate from internal detectors and third party vendors, there is no specific field that can be used to determine the alert category. Even though attempts are made to normalise alerts into a consistent format, often times the fields are missing or can semantically have different meanings. For example, the MITRE techniques reported could be used to categorise the alerts; however, these are often missing or incorrectly assigned.

A procedure was developed to generate an alert category which uses a combination of the detector and rule that generated the alert, when available, and the alert title. For alerts originating from internal systems, the combination of the rule and detector provide enough information to categorise an alert. For alerts ingested from third parties this information is not necessarily provided or too generic. In these cases the normalised alert title is used. Entities are often embedded in alert titles, e.g., ``Potential stolen user credential for user@domain''. Regular expressions are used to strip entities from within the title before being used for categorisation. 

After applying categorisation to the dataset, there were a total of $4,002$ categories over the 6-month period. Note that $2,246$ categories make up $99.8\%$ of the data. The alert objects can often have misconfigured inputs for the alert title or inconsistently formatted data, resulting in non-ideal categories being assigned in a small number of cases. In practice, these could be removed from the training data however this has very little effect on the results. It is assumed for the rest of the paper that an alert has an associated category $\mathcal{C} = \{c_1, \ldots, c_m\}$.

\section{Methodology}\label{sec:methodology}

 In Section~\ref{sec:alertdata}, a description of the alerts and the SOC workflow at \Secureworks{} was provided. However, the methodology presented is broadly applicable to any alerts triaged by an analyst, as it only requires that each alert has a category, entities, and the label or action assigned by the analyst.

A supervised ML approach is used over alerts that have been triaged by an analyst. First, if necessary, categorisation should be performed so that each alert belongs to a single category representing the attack behaviour.
 Feature extraction extracts static and dynamic features from the alert objects. These are used to train a classifier that predicts which action is taken by an analyst.

To deploy AACT in a live SOC environment, there would be two phases: training and scoring. In training, historical labelled alerts are used to train a model, using the static and dynamic features.

\begin{figure*}[ht]
   \centering
   \includesvg[width=0.95\textwidth]{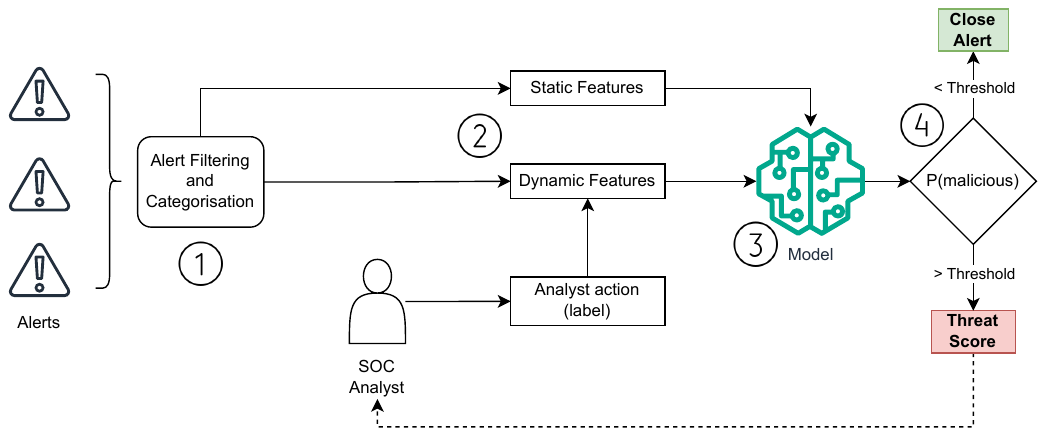}
   \caption{An overview of AACT}
   \label{fig:diagram}
\end{figure*}

Figure~\ref{fig:diagram} shows an overview of the live scoring phase. First any necessary filtering and enrichment's, such as categorisation, are applied to streaming alert data. Second the dynamic and static features are extracted from these alerts. Third, the model is applied, which produces a score for each finite set of actions that an analyst can take. In the final step, four, if the score for the alert being labelled malicious is below some threshold the alert could be automatically closed and removed from the analyst queue. The remaining scored alerts are then prioritised and sent to an analyst for review. Once the analyst has taken an action on the alert this is fed back into the module that calculates the dynamic features that are based on analyst actions. 

 The following subsections will explore each component in further detail.

\subsection{Feature Extraction}\label{sec:features}
One of the key components of the model is the assumption that the history of triage actions taken by analysts on alerts is a good predictor for what action will be taken in the future. 

The features aim to both measure the short-term and long-term trends for analyst actions performed on categories of alerts, as the interplay between these are important to accurately predict the outcome of future alerts. 

The recent history of analyst actions is important because it captures current and emerging trends about threats in a tenant's environment, enabling the system to learn and react in real-time. The long-term trends provide the historical context needed about overall false positive rates\footnote{A false positive in this sense means alert categories which tend to generate alerts that don't result in real security incidents.} for alert categories when viewed over a longer time frame, which would be missed by a shorter term view. In addition, if some categories are rare and have not been observed in the recent past, the historical data provides a baseline in lieu of more recent information. By combining trends across varying time windows, the system can form a robust model that results in more accurate predictions.

When analysts are triaging alerts, one consideration is how \emph{new} an alert is for the entities in the alert or for a tenant's environment. To encode this part of the decision-making process, features are generated that calculate how recent an alert category was seen for the tenant and for entities within the alert. 

The features described above are referred to as dynamic features as they change with respect to other alerts in the environment. In addition to these dynamic features, static features extracted from the alert itself are also used. These are both described in detail in the next sections.

In Sections~\ref{sec:dynamicfeatures} and \ref{sec:staticfeatures} the features are described for a general SOC workflow without specifying any specific triage action that has to be performed. Section~\ref{sec:secureworksfeatures} further discusses how these are modified and expanded on for the \Secureworks{} managed SOC workflow.

\subsubsection{Dynamic Features}\label{sec:dynamicfeatures}
Throughout this section, notationally, any dependence on a tenant is dropped. All features can be generated both for one tenant and across multiple tenants if applicable, see Section \ref{sec:secureworksfeatures} for further discussion.

Let $A^c(t, s)\in A=\{a_1, \ldots, a_n\}$ denote the categorical random variable for the action taken at time $s$ for an alert occurring at time $t$ of category $c$, e.g. nominate for further investigation, assign a label of malicious or benign. A list of counts for each action $a_i$ are generated over a lookback window $\delta$ for the number of occurrences of action $a_i$ for alerts of category $c$ over $[t-\delta, t)$,
\begin{equation}
\alpha^c_{a_i} = \sum_{\{A^c(t', s): t', s\in [t-\delta, t)\}} \mathbb{I}(A^c(t', s) = a_i).
\label{eq:category_count}
\end{equation}
Define a resolved alert as one that has been triaged and has had a corresponding triage action taken. Then, $\mathbb{I}(A^c(t', s) = a_i)$ denotes the indicator function taking the value $1$ if the action taken for an alert of category $c$ was $a_i$ and both the time the alert was created and the alert was resolved occurred within the lookback window $[t-\delta, t)$.  Note that in a slight abuse of notation, the dependency on the lookback window $\delta$ has been dropped. In practice, multiple lookback windows may be chosen. To choose which lookback windows to use, the correlation can be calculated over multiple $\delta$'s to maximise the variance across the lookback windows, see Section~\ref{sec:secureworksfeatures}. When generating local features per tenant, equation~\eqref{eq:category_count} would be calculated only over other alerts that occurred within the same tenant. For global features, it would be calculated across all alerts irrespective of the tenant.

It is desirable to normalise the counts so that they are comparable across different alert categories, some of which inherently generate more alerts than others, as well as smooth away any seasonality. The set of features then for an alert category is:
\begin{equation}
\mathcal{F}^c = \left\{f^c_{a_i} = \frac{\alpha^c_{a_i}}{\sum_{a\in A}\alpha^c_a}: i = 1\ldots n-1\right\}.
\label{eq:category_features}
\end{equation}
Note that the features are calculated over all but one of the triage actions $a_n$, say, as $f^c_{a_n}$ is perfectly correlated with $f^c_{a_1}$, \ldots, $f^c_{a_{n-1}}$.  This is equivalent to using as features an estimate, based on the data over the lookback window, of the probability of observing action $a_i$ for an alert given the category is $c$, $P(A(t) = a_i | C=c)$.

Let $X(t) = (\vectorentity{_1}, \ldots, \vectorentity{_k})$ denote the set of $k$ entities in an alert occurring at time $t$. Similarly, for each entity, a count for the number of occurrences of action $a_i$ for alerts that contain that entity over $[t-\delta, t)$ is obtained
\begin{equation}
\alpha^{\vectorentity{_j}}_{a_i} = \sum_{\{A(t',s): t',s\in [t-\delta, t)\}} \mathbb{I} (A(t',s) = a_i \cap \vectorentity{_j} \in X(t') ).
\label{eq:entity_counts}
\end{equation}
$\mathbb{I} (A(t',s) = a_i \cap \vectorentity{_j} \in X(t') )$ denotes the indicator function taking the value $1$ if the action taken for an alert is $a_i$, the entity $\vectorentity{_j}$ is contained in the list of entities, and both the time the alert was created and the triage action was taken occurred within the lookback window $[t-\delta, t)$. Note that for the entity features, there is no dependency on the alert category and the sum is over all alerts in the window. 

These are also normalised so that 
\begin{equation}
f^{x_j}_{a_i} = \frac{\alpha^{\vectorentity{_j}}_{a_i}}{\sum_{a\in A} \alpha^{\vectorentity{_j}}_{a_i}},
\label{eq:normalized_entity_ratio}
\end{equation}
which is equivalent to an estimate of the probability of seeing action $a_i$ for an alert given the entities within the alert contain $x_j$, $P(A(t) = a_i | x_j \in X)$.

In order to generate a fixed number of features per alert, summary statistics such as the mean or max can be used over the set of normalized entity counts for each action $\{f^{x_j}_{a_i}: \vectorentity{_j} \in X(t)\}$.

The set of features for the alert entities based on the max counts, say, are:
\begin{equation}
\mathcal{F^X} = \left\{f^{X}_{a_i} = \max_{j\in k}f^{x_j}_{a_i}:  i = 1\ldots n\right\}.
\label{eq:entity_features}
\end{equation}

For some alerts the denominator can be $0$ for $f^c_{a_i}$ or $f^{\vectorentity{_j}}_{a_i}$ if the category is new for the tenant or if the entity has not been alerted on in the time window, respectively. In these cases, $f^c_{a_i}$ and $f^{\vectorentity{_j}}_{a_i}$ are set to $0$. An alternative more principled approach could be to place a prior on the number of occurrences of an action in a Bayesian fashion.

 When generating the category and entity features for an alert occurring at $t$, any alerts occurring in $[t-\delta, t)$ that have not been triaged and resolved are not included in the denominator for equations \eqref{eq:category_features} and \eqref{eq:normalized_entity_ratio}. Two final additional dynamic features are included: the total number of overall alerts and the percentage of those that have been resolved both for the category and entity features, denoted $\mathcal{F^{C,R}}$ and $\mathcal{F^{X,R}}$, respectively. These additional features are particularly useful for balancing situations where there is a high proportion of alerts associated with action $a_i$, yet only a few alerts have been resolved, compared to scenarios with a large number of resolved alerts. Typically, these features only need to be included for shorter look-back windows $\delta$, since over longer time frames, most or all alerts will have been resolved.

Finally, features are defined that capture how ``new'' an alert is for a tenant's environment.

First is the time last seen for an alert occurring at $t$ of category $c$
\begin{equation}
f^c_t = t - \max(\{t' < t: C(t') = c\}),
\label{eq:time_last_seen_category}
\end{equation}
where $C(t)\in\mathcal{C}$ denotes the category of an alert.

To capture how new the alert category is for an entity $\vectorentity{_j}$, define
\begin{equation}
f^{c,\vectorentity{_j}}_{t} = t - \max(\{t' < t: C(t') = c, \vectorentity{_j} \in X(t')\}).
\label{eq:time_last_seen_entity}
\end{equation}
Similarly to above, to have a fixed number of features for each alert, summary statistics can be taken over $\{f^{c,\vectorentity{_j}}_{t}: \vectorentity{_j} \in X(t)\}$ denoted $f^{c,X}_t$. In practice, these rarity features will be left-censored. To account for this, an initial amount of data over the training period should be kept back to calculate equations \eqref{eq:time_last_seen_category} and \eqref{eq:time_last_seen_entity}.

For each alert, the complete set of dynamic features extracted are $\mathcal{F}=\{\mathcal{F}^c, \mathcal{F^{C,R}}, \mathcal{F}^X, \mathcal{F^{X,R}},  f^c_t, f^{c,X}_t\}$.

By encoding features that learn from the analysts triage actions using alert similarity, the model is not fitting on specific alert attributes whose relationship with the dependent variable may vary rapidly over time. For example, there are times whereby a specific alert entity may be involved in a real security incident and other times where alerts originating from that same entity are benign. If this entity was explicitly encoded as a feature (using one-hot encoding for example) frequent retraining of the model would be required to adapt to the current security state of that entity. 

The features are also interpretable by SOC analysts. It is easy to understand how the score for an alert is affected by a) previous triage actions on similar alerts and b) how unlikely it is to see the category of an alert for the tenant or entity. Based on discussions with subject matter experts (SME), both of these statistics are key parts of what an analyst looks at when they triage an alert. 

For real-time scoring of alerts, the primary complexity of the model is in maintaining up-to-date counts of analyst triage actions over alert categories, tenants and entities. This is required to compute the dynamic features using the most recent information and avoid performance degradation.

\subsubsection{Static Features}\label{sec:staticfeatures}
While previous features aim at encoding dynamic alert trends, several static features are also added that can directly be extracted from the alert object irrespective of any other alerts. To avoid the need for frequent retraining, these features should not have a relationship that changes quickly over time with the analyst action that is being predicted. Section~\ref{sec:secureworksfeatures} lists the static features extracted for the \Secureworks{} alert data.

\subsection{\Secureworks{}}\label{sec:secureworksfeatures}
A description is now given on how the features are generated for the \Secureworks{} SOC workflow.  As discussed in Section~\ref{sec:alertdata}, the goal is to predict if an alert occurring at time $t$ is added to an investigation and a notification sent to the client of malicious activity occurring in their environment.

\subsubsection{Dynamic Features}
For equations \eqref{eq:category_features} and \eqref{eq:entity_features}, the actions an analyst can take are $A^c(t) = (a^c_1, a^c_2)$, where $a^c_1$ corresponds to an alert getting added to an investigation and conversely $a^c_2$ is not added. In addition to these features, it is desirable to capture the label applied by the analyst to the alert, see Section~\ref{sec:alertdata}. 
Two extra sets of features are generated as in \eqref{eq:category_features} and \eqref{eq:entity_features}, where now the set of actions $L^c(t) = (l^c_1, l^c_2)$ is if a label was applied to the alert indicating that it was malicious, $l^c_1$, or benign $l^c_2$. These extra sets of features will be denoted $\mathcal{F}^c_{L}$ and $\mathcal{F}^X_{L}$.

As \Secureworks{} is a MDR provider, alerts from multiple tenants environments are triaged. For this reason, the features for the alert categories, $\mathcal{F}^c$ and $\mathcal{F}^c_{L}$ are computed per tenant environment and then across all tenant environments, referred to as the global features. This tailors the model per tenant while learning from global trends for categories of alerts. 

\begin{figure}[h]
    \centering
    \includegraphics[width=1\columnwidth]{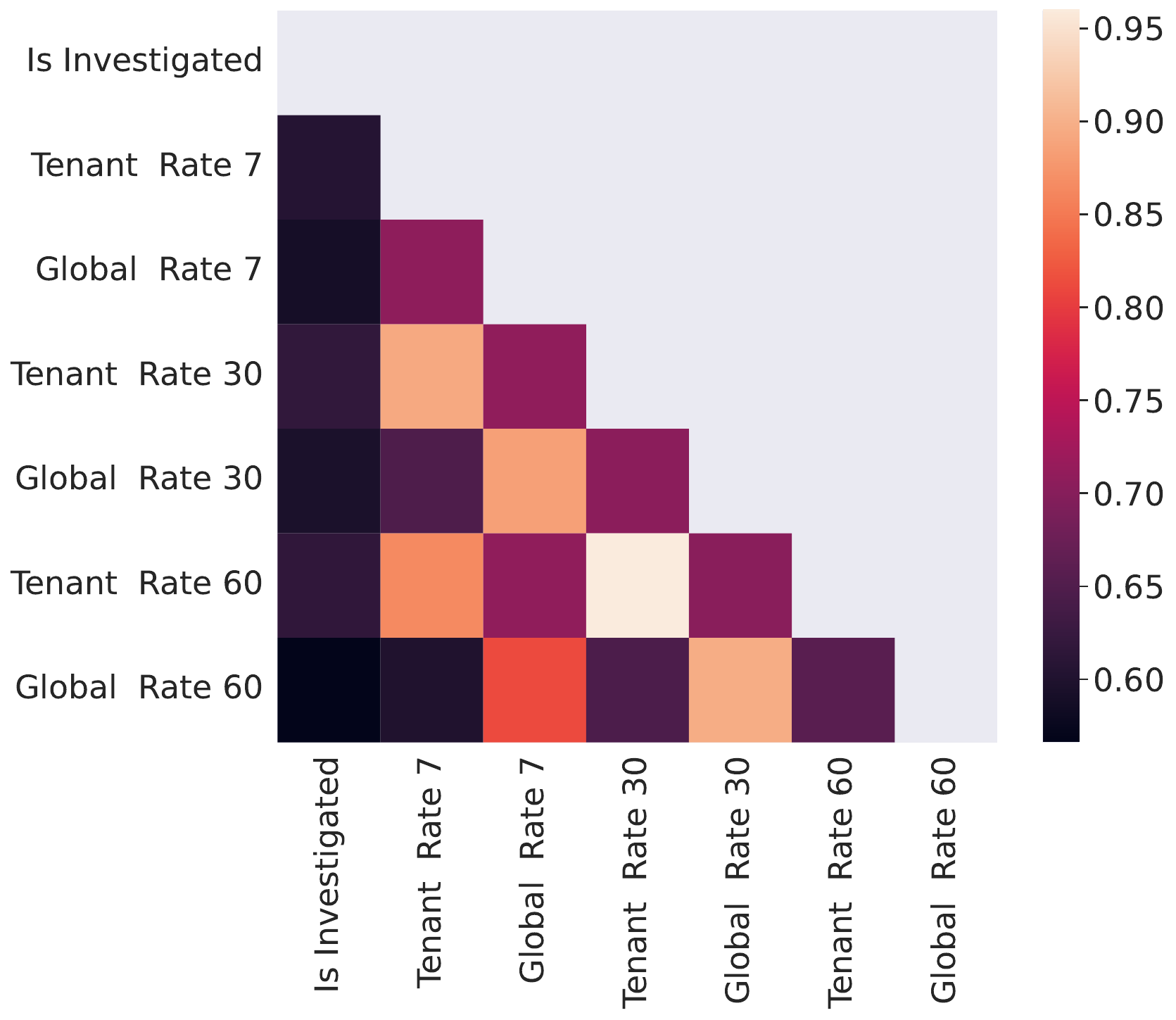}
    \caption{Correlation matrix for the `is investigated' dependent variable and the category investigation rates with lookback windows of $7$, $30$ and $60$ days.}
    \label{fig:correlation_matrix}
\end{figure}

The tenant, global and entity features for the investigation and malicious label rates , $\mathcal{F}^c$, $\mathcal{F}^X$, $\mathcal{F}^c_{L}$ and $\mathcal{F}^X_{L}$  were computed over a lookback window of $\delta = 1$ day, $\delta = 7$ days, $\delta = 30$ days and $\delta = 60$ days. Figure~\ref{fig:correlation_matrix} shows the lower triangular correlation matrix for $\mathcal{F}^c$ over the longer lookback windows for the tenant and global features and the dependent variable indicating whether the alert was added to an investigation. There is a strong correlation between the 30 day and 60 day lookback windows, achieving a correlation of upwards of 0.8. For this reason, the lookback window of 60 days is dropped from the set of features. It can also be seen from Figure~\ref{fig:correlation_matrix} that the investigation rates over the various time windows for the alert categories are correlated with an alert being investigated. For the features for the number of alerts and the resolved ratios, $\mathcal{F^{C,R}}$ and $\mathcal{F^{X,R}}$, they are computed only over the one day lookback, as beyond that the resolved ratio tends to $1$. Here, resolved is defined as either having been added to an investigation or labelled and closed. 

For the entity features $\mathcal{F}^X$ and $\mathcal{F}^X_{L}$,  the max is taken over the list of entities within the alert, see equation \eqref{eq:entity_features}. After discussing with SMEs, it was suggested that if there was any key entity in the alert that had a recent high investigation or malicious label rate, then it should be prioritised for investigation by an analyst. For the features related to when the alert category was last seen for the entity, equation \eqref{eq:time_last_seen_entity}, the maximum over the set of entities is also taken.

\subsubsection{Static Features}
The following are the static features extracted from the alerts. First is the number of entities associated with an alert, as well as the number of entity relationships, i.e., the actions between entities, e.g., \emph{user A} authenticating to \emph{service Z}. All alerts that are reported by the \Secureworks{} platform have one or more corresponding MITRE tactics and techniques that are associated with the alert. MITRE proposes a framework detailing attacker behaviour~\cite{strom2018mitre}, specifying 1) Tactics, representing the steps they consider in an attack chain, and 2) Techniques, which identifies actions attackers take as part of these tactics. From these, two features are constructed: the number of techniques associated with the alert, along with the farthest tactic on the attack chain, represented by a number between 0 and 14, where 0 signifies no tactic was present. 

Table~\ref{tab:features} summarises the complete feature set extracted for each alert.
\begin{table}
\centering
\begin{tabular}{ |l| } 
\hline
{\bf Dynamic Features} \\
\hline
Max investigation rate for entities ($\delta = 1, 7, 30$ days) \\ 
Max malicious label rate for entities ($\delta = 1, 7, 30$ days) \\ 
Max resolved rate for entities ($\delta = 1$ day) \\ 
Category investigation rate per tenant ($\delta= 1, 7, 30$ days) \\ 
Category investigation rate global ($\delta= 1, 7, 30$ days)\\ 
Category malicious label rate per tenant ($\delta= 1, 7, 30$ days)\\
Category malicious label rate global ($\delta= 1, 7, 30$ days)\\
Category resolved rate per tenant ($\delta=1$ day)\\
Category resolved rate global ($\delta=1$ day)\\
Category total alerts per tenant ($\delta=1$ day)\\
Category total alerts global ($\delta=1$ day)\\
Delta since category was last seen for tenant \\ 
Max delta since category was last seen for entity\\ 
\hline
{\bf Static Features}\\
\hline
Entities count \\ 
Entities relationship count \\
Max tactic score \\
Tactic count \\
\hline
\end{tabular}
\caption{Complete set of features extracted per alert}
\label{tab:features}
\end{table}

\subsection{Classifier}\label{sec:classifier}
A classifier can be trained over the possible set of triage actions, $A$, an analyst can take using each of the features discussed in Section~\ref{sec:features}. If there are multiple actions that are being predicted, a multi-class classifier, such as a decision tree, is appropriate. If there are only two actions, a binary classifier like logistic regression can be used. For both examples presented in Section~\ref{sec:results} a gradient boosting tree was used as that provided the best performance. To evaluate the model, cross-validation with a time-series split should be used. This type of evaluation, which always tests folds on newer data, is crucial when evaluating time-series data, especially data stemming from cybersecurity applications. The features generated rely on knowing counts of previous actions taken on alerts: using classical cross-validation techniques would provide training samples with future information it shouldn't have. In addition, cyber attacks evolve and adapt over time, and simply using randomised folds would leak future attack trends to the training set and inflate the model's performance. The time-series split for evaluation is also representative of how the model would perform in a live environment with retraining.

\section{Evaluation}\label{sec:results}
To evaluate AACT, we used two datasets: the publicly available AIT Alert Data Set \cite{landauer2024} and an internal \Secureworks{} alert dataset, described in Section~\ref{sec:alertdata}. Data preparation and feature extraction was implemented using Spark~\footnote{https://spark.apache.org/docs/latest/api/python/index.html}, while Scikit-learn~\footnote{https://scikit-learn.org/stable/} was used for model training and evaluation. 

In both datasets, analysts have two possible actions indicating an alert as either malicious or benign. 
This evaluation implements a gradient boosting classifier, comparing its performance against other models on the \Secureworks{} dataset. The priority score is simply the score ranging from $[0, 1]$ output by the gradient boosting classifier.  Standard performance metrics such as Accuracy, Precision, Recall, F1-Score, and ROC AUC are used for evaluation. 
Analysts often close or resolve alerts unrelated to real security incidents. Triaging each alert is time-intensive, driving up resource costs significantly. By accurately predicting and removing benign alerts before they reach analysts, the system can greatly reduce workload and resource use. Thus, if there is high enough confidence that an alert is malicious, the system can close the alert before it ever reaches an analyst. Consequently, understanding the model's effectiveness in reducing the volume of alerts analysts must review, while minimizing incorrectly closed alerts, is crucial. 
Falsely closed alerts are critical since they might reduce the visibility into an on-going attack. Figure~\ref{fig:diagram} illustrates how alerts below a defined priority threshold are removed from the analyst queue, reducing the volume of alerts requiring review. Therefore, the final metric used is alert reduction, which is defined as the proportion of alerts eliminated from the queue. 

Note that there is a large difference in the degree of alert reduction between the internal alert dataset and the AIT dataset and, more broadly, those found in the literature, such as \cite{landauer2022, hassan2020, deepcase}. This difference is due to the fact that the AIT alert data and the datasets used in the literature often consist of low-level alerts from IDS systems, lacking filtering or de-duplication, leading to a higher level of alert reduction, contrary to our internal dataset.

AACT is compared with a baseline and the DeepCase algorithm \cite{deepcase}. However, the DeepCase comparison is limited to the AIT alert dataset, as it was unsuitable for the \Secureworks{} dataset due to its approach of grouping alerts by single entities. Handling alerts with multiple entities of various types, such as hosts, users, and processes, would require further research. Despite applying a few approaches to tackle the multi-entity issue, DeepCase was still less effective as it was not designed to handle alerts originating from a pipeline of alerts that have already gone through many layers of reduction and correlation. The method’s strength lies in using low-level contextual information to group, cluster, and prioritize alerts, a feature absent in the available \Secureworks{} data, resulting in suboptimal performance.

\subsection{\Secureworks{} dataset}\label{sec:results_secureworks}

As discussed in Sections~\ref{sec:alertdata} and \ref{sec:secureworksfeatures} the binary state of whether an alert gets added to an investigation is the analyst action being predicted for this example.

The alert data, described in Section~\ref{sec:alertdata} is a subset of those processed by the \Secureworks{} security platform. There are \totalalertstesttenantsremoved{} alerts over a 6-month period.
As the algorithm described in this paper has been running in a live environment automating the resolution and closing of alerts, a final filter is applied to remove those alerts that were closed by this automation in the past, to avoid a direct feedback loop, see also Section~\ref{sec:live-implementation}. The final number of alerts available as training samples following this is \trainingsamplesprioresolvedremoved{}. Of these, \trainingsamplespositiveclass{} were added to an investigation, i.e., are from the positive class for the binary classification problem. 

The features used are those described in Section~\ref{sec:secureworksfeatures}. For the feature extraction, an additional one month of data prior to the training data is used to calculate the one-month investigation rates for the alert category, equation \eqref{eq:category_features}, and the delta since the category was last seen for the tenant and entities, equations \eqref{eq:time_last_seen_category} and \eqref{eq:time_last_seen_entity}. Multiple algorithms were explored to train and evaluate the model, namely Logistic Regression, Random Forest classification and Gradient Boosting classification. The average of each algorithm's performance metrics obtained from the 10-fold cross-validation over time series splits are shown in Table~\ref{tab:results} for a 0.5 threshold.
Gradient Boosting classification obtained a slightly better performance and is used in this approach. Note that the Random Forest model had better precision at the cost of a lower recall. In our case, missing true critical alerts has a larger negative impact to the business than incorrectly marking alerts as critical. For this reason, we use the Gradient Boosting classifier which obtains better recall, hence fewer false negatives. 

For comparison, we use the global investigation ratio of alert categories over 30 days to represent the baseline model. Thresholds are simply set on the investigation ratio itself to calculate the metrics for the baseline, with no classification involved.

\begin{figure}[b]
   \centering
   \includegraphics[width=1.0\columnwidth]{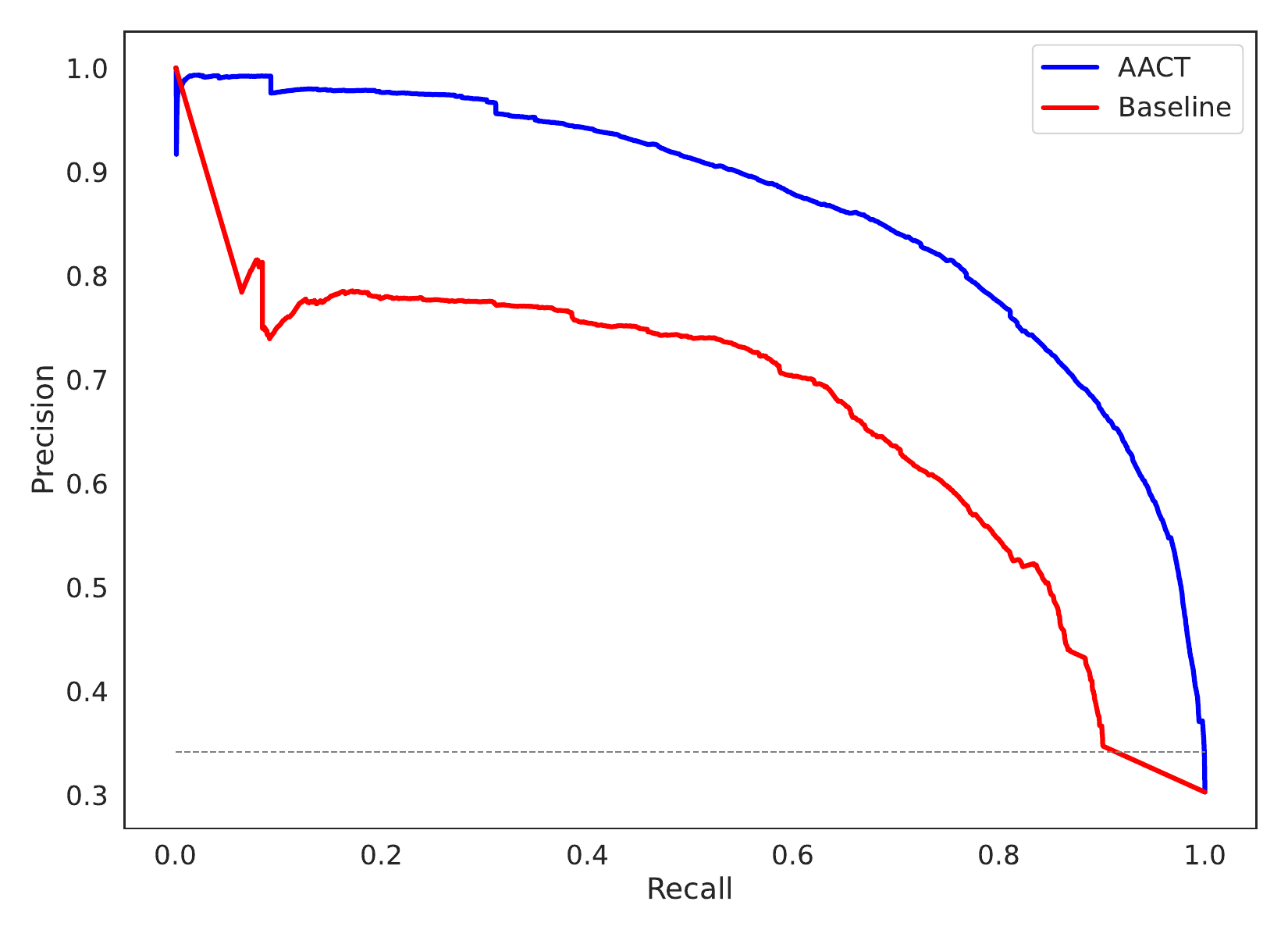}
   \caption{The precision versus recall curve at varying thresholds}
   \label{fig:precision_recall}
\end{figure}

\begin{figure}[h]
    \centering
    \includegraphics[width=1.0\columnwidth]{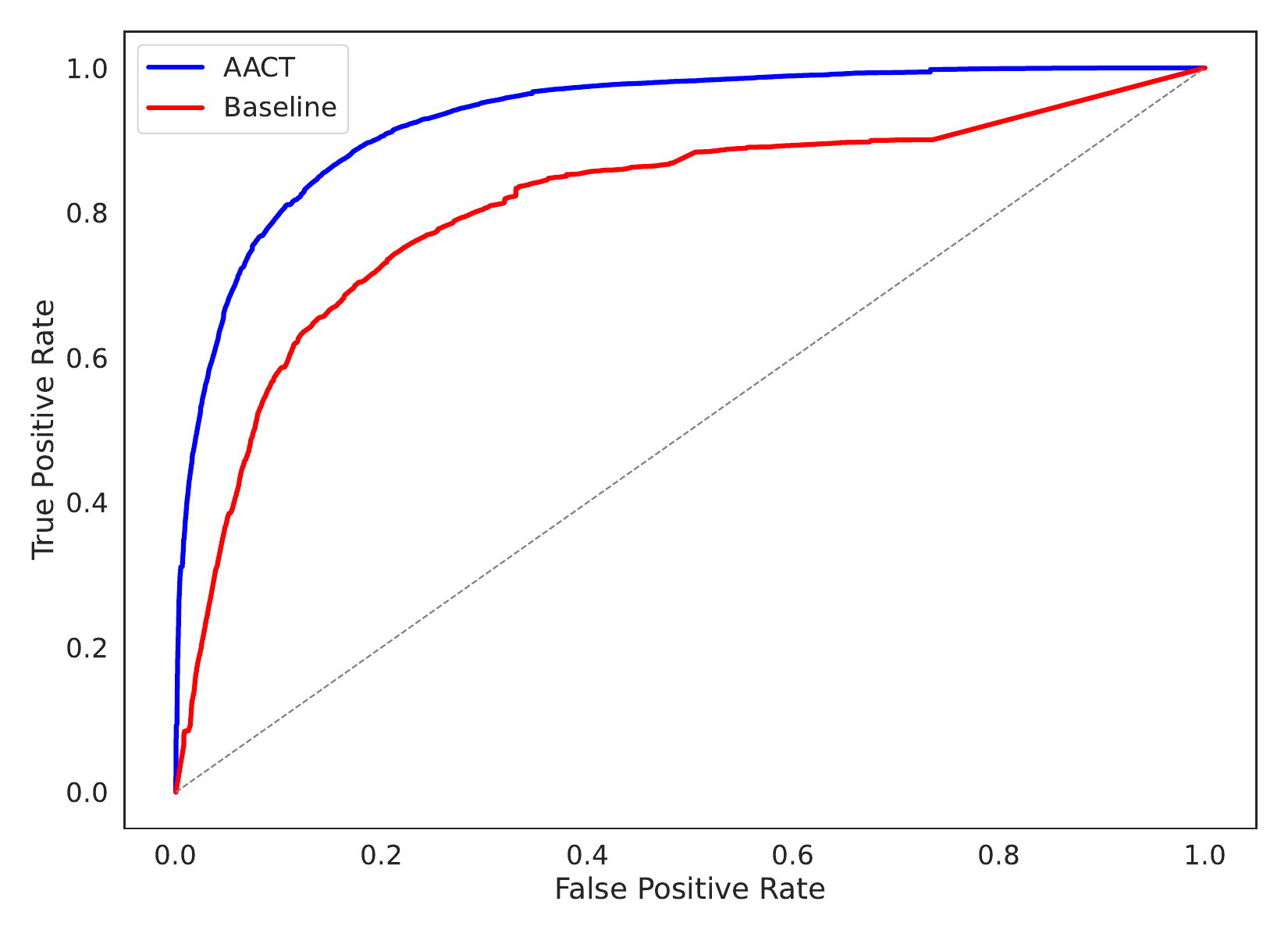}
    \caption{The ROC curve showing the false positive rate versus the true positive rate at varying thresholds}
    \label{fig:fpr_tpr}
\end{figure}

Figures~\ref{fig:precision_recall} and~\ref{fig:fpr_tpr} show the precision-recall and ROC curves over all folds. The results show good performance in predicting if an alert will be added to an investigation. Maximising recall (true positive rate) is important as misclassifying alerts that would have been added to an investigation can cause real security incidents to be missed. Figure~\ref{fig:fpr_tpr} shows that a threshold can be chosen that would achieve a high true positive rate while maintaining a low false positive rate. The figures show that AACT outperforms the baseline, and bring additional value over simply increasing the priority of alert categories that have a high investigation rate.

\begin{table*}[t]
  \begin{center}
    \caption{Model evaluation performance}
    \begin{tabular}{|l|c|c|c|c|r|}
      \hline
        \textbf{Algorithm} & \textbf{Accuracy} & \textbf{Precision} & \textbf{Recall} & \textbf{F1-Score} & \textbf{ROC AUC} \\
      \hline
        Gradient Boosting Classifier & 87.02 & 78.79 & 78.21 & 78.27 & 93.32 \\
        Random Forest Classifier & 87.73 & 82.11 & 74.97 & 78.22 & 93.88 \\
        Logistic Regression Classifier & 85.44 & 76.77 & 74.08 & 75.10 & 91.86 \\
      \hline
    \end{tabular}
    \label{tab:results}
  \end{center}
\end{table*}

\begin{figure}[h]
    \centering
    \includegraphics[width=1.0\columnwidth]{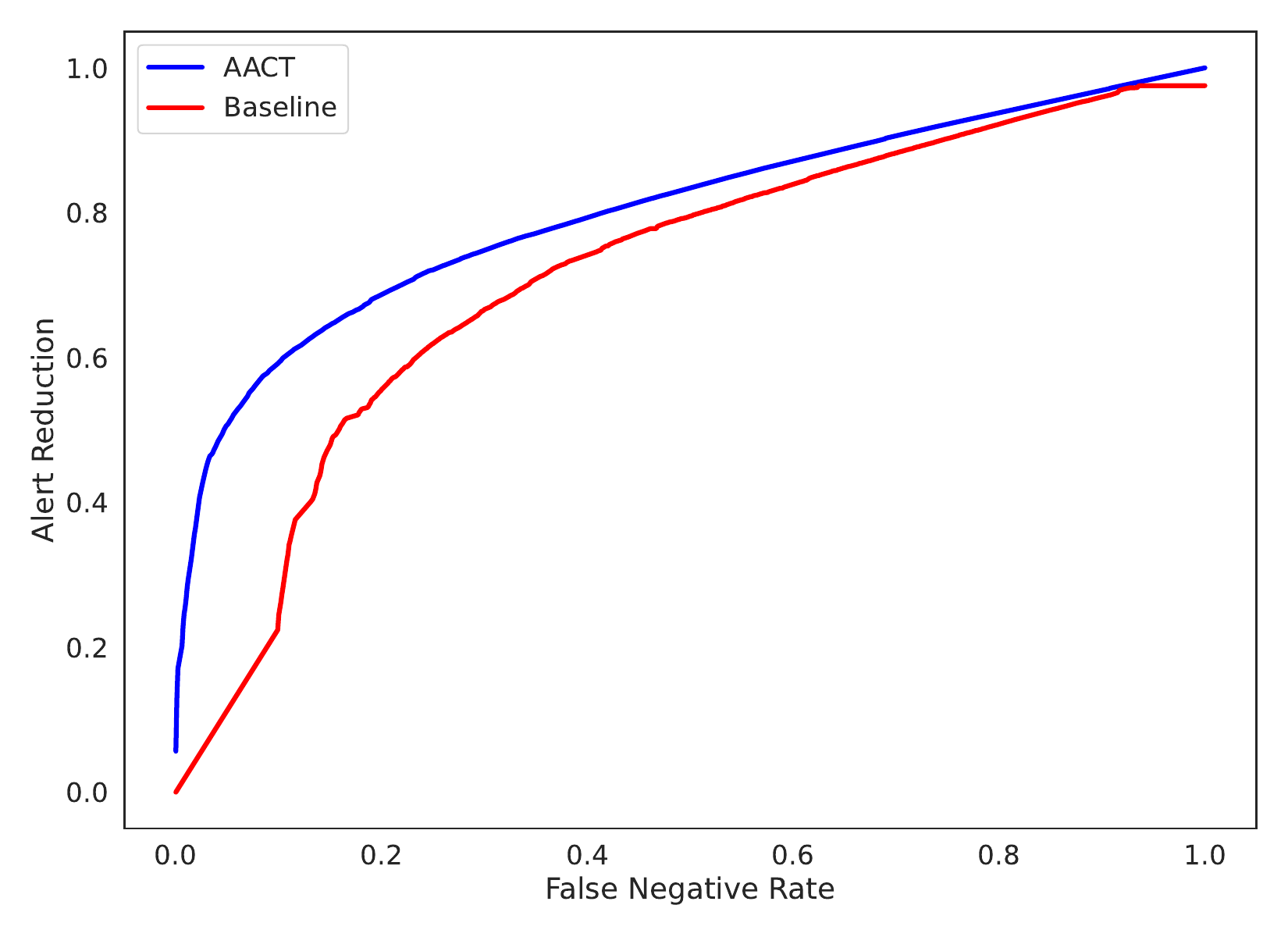}
    \caption{The percentage of alerts removed from the analyst queue vs the false negative rate at varying thresholds}
    \label{fig:dropped_alerts}
\end{figure}

Figure~\ref{fig:dropped_alerts} shows the alert reduction versus the false negative rate over varying thresholds. In this context, a false negative is an alert that should have been investigated, but was closed incorrectly as benign. As can be seen, a threshold can be chosen such that $34\%$ of alerts can be removed from an analysts' queue with a false negative rate of $1.5\%$. Note that if you include the $728,834$ alerts removed from the analysis due to them being closed by AACT, the reduction rate is closer to $74\%$. AACT does not train on its own output, to prevent a direct feedback loop. Section~\ref{sec:live-implementation} will give further details into this process, through a description of the live model and its performance.

\begin{figure}[h]
    \centering
    \includegraphics[width=1.0\columnwidth]{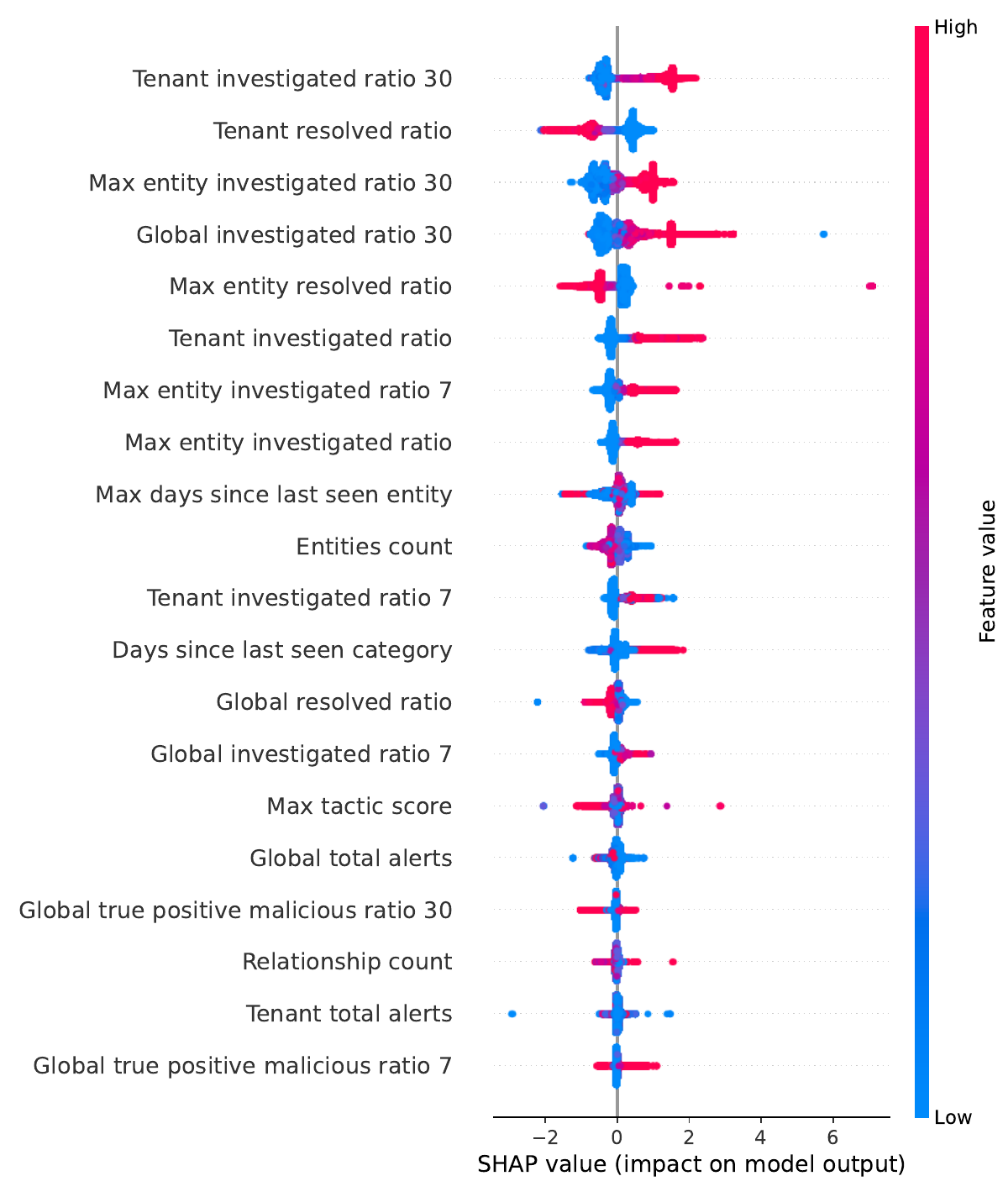}
    \caption{Shapley values for the model's top features}
    \label{fig:top_features}
\end{figure}

To identify the key features driving the predictions of the model, the Shapley values~\cite{winter2002shapley} are computed and shown in Figure~\ref{fig:top_features}. The alert category investigation rate for a tenant over all lookback windows has the largest impact on the model's decision, with high alert category investigation rates increasing the probability that future alerts will be added to an investigation. This is supported by the observation that even the simple baseline using the 30-day investigation rate shows reasonable performance. Overall, tenant-specific features have a larger impact on the prediction than global features, indicating that the model adapts well to the specific tenant environments.

\subsection{AIT Open Alert dataset}
The Austrian Institute of Technology (AIT) Log Data Set \cite{landauer2023, landauer2021} includes system and network logs from eight testbed networks, which were subjected to various multi-step attack scenarios over a three week simulation period, from January 14 to February 2, 2022. The AIT alert dataset \cite{landauer2024} comprises alerts generated from these logs using three intrusion detection systems: Suricata, Wazuh, and AMiner, totaling 2,655,821 alerts in JSON format normalized across the different detectors \footnote{\url{https://github.com/ait-aecid/alert-data-set}}. 
The normalized alerts serve as input for this example, with each record containing fields such as timestamp, name (corresponding to the detector and rule that generated the alert), host, time label, and event label. The event label shows if an alert is linked to an attack-related log event, while the time label indicates if it occurred during an attack scenario and indicates false positive or attack type. Since benign alerts can occur during attacks, the event label is used if the event happened during the attack window. Notably, some events are incorrectly labelled as malicious outside the attack window and are marked as benign.

In this simulated scenario, we assume that alerts are triaged during standard SOC operations. As a result of this triage process, alerts are classified as either {\it malicious}, if the event label indicates an attack, or {\it benign}. For the dynamic features, as described in equations \ref{eq:category_features} and \ref{eq:entity_features} in Section~\ref{sec:dynamicfeatures}, $A_c(t) = (a_1^c, a_2^c)$, where $a_1^c$ represents an alert labelled as malicious and $a_2^c$ represents an alert labelled as benign. The alert category is simply the name, as no entities are embedded in the name. No further categorization is needed. There are 93 categories in total across all alerts. Given the delay between when an alert is received and when an analyst labels it, it is assumed that the labelling time is uniformly distributed over a range of 1 to 16 minutes after the alert is created. This assumption is based on the 5th and 95th percentiles of triage times for alerts at \Secureworks{}. Including this additional timestamp is crucial, because the features are generated based on these analyst labels, as discussed in Section~\ref{sec:dynamicfeatures}. Assuming that the label existed at the time of alert creation could artificially enhance the model's performance and would not accurately reflect real SOC operations. Each test bed network is treated as a tenant. The complete set of dynamic features extracted from this dataset is provided in Table~\ref{tab:aitfeatures}. Since only one entity is reported per alert, calculating the maximum over the list of entities is unnecessary. Longer time frames for resolved and malicious label rates are not used due to the data covering only approximately three weeks.

\begin{table}
\centering
\begin{tabular}{ |l| } 
\hline
{\bf Dynamic Features} \\
\hline
Entity resolved rate ($\delta = 1$ day) \\ 
Entity malicious label rate ($\delta = 1$ day) \\ 
Category malicious label rate per tenant ($\delta= 1$ days)\\
Category malicious label rate global ($\delta= 1$ days)\\
Category resolved rate per tenant ($\delta=1$ day)\\
Category resolved rate global ($\delta=1$ day)\\
Category total alerts per tenant ($\delta=1$ day)\\
Category total alerts global ($\delta=1$ day)\\
Delta since category was last seen for tenant \\ 
Delta since category was last seen for entity\\ 
\hline
\end{tabular}
\caption{Complete set of features extracted per alert in the AIT dataset.}
\label{tab:aitfeatures}
\end{table}

The data from all tenants is combined and arranged in chronological order. Due to the data size, only two-fold time-series cross-validation is employed in order to ensure a sufficient number of malicious samples in each fold. Default parameters were used for the gradient boosting classifier.

\subsubsection{A comparison between AACT and DeepCase}
The methodology is compared against DeepCase, \cite{deepcase} using the code provided by \url{https://deepcase.readthedocs.io}. DeepCase initially correlates alerts (referred to as security events in \cite{deepcase}) using the Context Builder, which leverages a recurrent neural network with an attention mechanism to analyze sequences of alerts that occur in close temporal proximity on the same host. It then uses an Interpreter to compare all correlated alerts and group them into similar clusters. These clusters are subsequently presented to a security operator for labelling during a manual analysis phase. Following this, DeepCase operates in a semi-automatic mode, where the Interpreter can compare new alert sequences to existing clusters using the attention vectors generated by the Context Builder for the new sequence. If the new sequence matches known benign clusters, they can be automatically closed without human analyst intervention.

\begin{table*}[h!]
  \begin{center}
    \caption{Model evaluation performance}
    \begin{tabular}{|l|c|c|c|c|c|c|}
      \hline
        \textbf{Method}  & \textbf{Precision} & \textbf{Recall} & \textbf{F1-Score} & \textbf{Alert Reduction} & \textbf{FNR}\\ 
      \hline
        AACT & 99.45 & 90.97 & 94.95 & 95.45 & 9.03 \\
        DeepCase & 97.25 & 66.97 & 79.24 & 89.74 & 33.03 \\
        Baseline & 96.70 & 87.55 & 91.90 & 95.45 & 12.44 \\
      \hline
    \end{tabular}
    \label{tab:results_ait}
  \end{center}
\end{table*}

For the comparison with AACT, the training data for each time-series split represents the manual analysis phase, while the test data represents the semi-automatic phase. The vocabulary size for the neural network is set to the number of unique categories, with all other parameters kept at the defaults used in \cite{deepcase}. During the manual phase, the maximum label of any individual alert is applied to the entire cluster. Using the average or minimum of all labels resulted in similar or worse performance. In the semi-automatic phase, each alert is assigned a score indicating whether it is benign, malicious or it is unscored if the Context Builder lacked confidence for a prediction, or if the nearest cluster was beyond a specified "epsilon" distance from the nearest sequence.

The average performance metrics for AACT and DeepCase across each fold are presented in Table~\ref{tab:results_ait} along with the baseline performance. For AACT, metrics are reported using a 0.5 threshold, where alerts with scores greater than or equal to 0.5 are considered malicious, otherwise benign. For DeepCase, results are further averaged over 10 runs. As a baseline, the global category malicious label rate is used similarly to Section~\ref{sec:results_secureworks}. The baseline threshold was chosen such that the Alert Reduction was the same as for AACT.

Table~\ref{tab:results_ait} shows that AACT outperforms other methods on this public dataset, achieving greater alert reduction and a lower FNR. The baseline also shows good performance with only a slightly higher FNR than AACT. The alerts generated from this synthetic dataset are significantly less noisy and diverse, originating from only three different traditional IDS systems, unlike the more complex data set from \Secureworks{}. As a result, even a simple baseline performs very well. It's important to note that the precision and recall for DeepCase are computed across all alerts, including those not labelled by the algorithm, to ensure a fair comparison with AACT. Out of the 127,152 alerts in the test set for each fold, an average of 8,558 alerts were not scored by DeepCase. Among these, 6,389 belonged to the positive class, indicating that many alerts where the Context Builder lacked confidence or were not sufficiently close to an existing cluster were more likely to be malicious than benign. If these unscored alerts are excluded as in \cite{deepcase}, DeepCase achieves a precision of $97.25\%$ and a recall of $93.2\%$.

DeepCase's comparatively lower performance may result from its inability to incorporate tenant-specific knowledge, treating alerts from all tenants as though they originate from the same source. This approach limits its ability to adapt to and learn from the unique environments of different tenants. Building separate models for each tenant was not feasible due to insufficient data.
Despite this, it is worth noting that DeepCase also clusters alerts through the Interpreter, allowing analysts to examine a subset of alert sequences within a cluster collectively, rather than triaging each alert individually. Among the unscored alerts, 6,725 formed 99 new clusters, which can be seen as an additional reduction of $4.5\%$, based on analysts reviewing 10 sequences per cluster before applying a label, as detailed in \cite{deepcase}. This clustering capability enhances efficiency and provides a valuable tool for security operations.

\section{Live Implementation}\label{sec:live-implementation}
The approach presented in this paper has been deployed in a production environment. The model has been used to de-escalate alerts for security analysts, drastically reducing the number of alerts presented to them on a daily basis. The de-escalation threshold is chosen based on what is an acceptable false negative rate to the user: a higher threshold will reduce more alerts, at the cost of missing more true incidents by closing legitimate alerts. 
 
When training future models, alerts that were closed by the automated live system are removed from the training data. Re-training the model runs in $\sim26$ minutes, using spark for distributed computing and a single large machine for model training. The model is then published as an endpoint and used to score all incoming alerts, closing alerts whose predicted probability of getting added to an investigation falls below a chosen threshold.  In the live implementation, AACT is a containerized application deployed via a Kubernetes\footnote{https://kubernetes.io/} cluster. A timeseries database is used to store counts of analyst actions over categories of alerts with a maximum of 5 minutes latency. The application is able to score 300 alerts per minute, with a latency ranging from 1.61 to 4.02 seconds, averaging 3.04 seconds with two containers which can be trivially scaled horizontally.

An important element to consider when deploying a predictive model in a live environment is to prevent a direct feedback loop \cite{sculley2015hidden}, where the model learns from its own predictions. The approach presented here learns from analyst labels applied to alerts and closes alerts before they reach a security analysts' work queue. To prevent the model from learning from its own predictions, and to be able to continuously assess performance of the model, a subset of alerts that would have otherwise been closed are sampled and sent to a human to process.  The sampling mechanism used is disproportionate stratified sampling~\cite{daniel2011sampling}, with the strata corresponding to the alert category. This is done in such a way that an equal proportion of each alert category is represented in the total sample. This provides diverse training samples for the future and continuous monitoring of the performance of the model via dashboards. 

The live implementation has resulted in a 61\% alert reduction over a 6-month period, with a false negative rate (FNR) of 1.36\%. This FNR is determined by the sampling mechanism discussed above. An FNR of 1.36\%, split over all clients, amounts to under 1 incorrectly closed alert per 
week per tenant over their entire network. Over the 6-month period, in all but two cases, the underlying security events linked to these missed alerts produced other alerts that were correctly predicted; hence the missed security incident FNR is lower than the model’s FNR. The missed events were reported post-hoc and corrected. Both were a result of mislabeled alerts from analysts, which is a limitation of the system, see Section~\ref{sec:discussion} for further discussion. In the majority of cases, mislabeled alerts are reported by the customer and corrected, so that by the time the model is trained, the labels are correct. This FNR was deemed acceptable by subject matter experts, given that the risk of human analysts mislabelling alerts increases as the number of alerts to triage increases.
\begin{figure}[h]
    \centering
    \includegraphics[width=1.0\columnwidth]{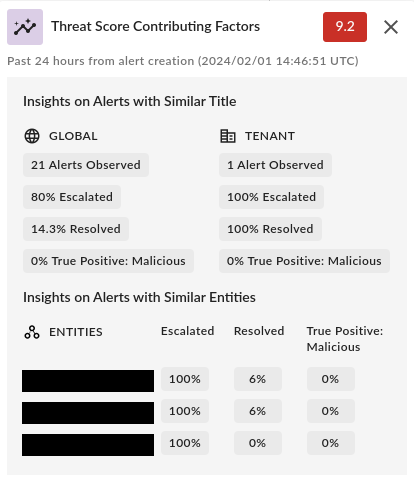}
    \caption{Threat Score display for SOC analysts}
    \label{fig:threat_score}
\end{figure}

A goal in deploying this framework is to create trust in the system by providing analysts with detailed insights into why an alert has been automatically closed, or its priority increased. 
Therefore, the alerts user interface (UI) in the framework shows the probability output by the model, referred to as the threat score, alongside the top impactful features behind the score.
The threat score shown is normalised to a score in the range $[0-10]$. Figure~\ref{fig:threat_score} depicts an example alert with a threat score of 9.2, with the features driving the model prediction shown under the score. In this example, the alert was given a high score since a high percentage of alerts with a similar category were investigated both globally and for that tenant. There were also a large percentage of investigations related to the entities in the alert.

While the system is running in the live environment, there are direct lines of communication with SOC analysts where issues can be reported. We constantly iterate on feedback improving the features and the way that results are shown in the UI. Metrics used to track SOC efficiency such as time to notify have been noticeably reduced by the deployment of AACT.

\section{Discussion}\label{sec:discussion}

A limitation of the approach is that mislabelling by human analysts can have a significant impact on the model as it is directly learning from their labels. During training on historical data, this is not a significant issue as mislabelled data is most often reported and corrected after the fact. However, it can have an effect during the live scoring phase when calculating the dynamic features, resulting in misclassified alerts. Future efforts to mitigate this is to weight analyst labels based on their experience level. Additionally, weights could be applied to the counts in \eqref{eq:category_count} and \eqref{eq:entity_counts} if there is a greater entropy in the number of analysts who have taken that action. For example, if multiple analysts have previously labelled an alert category as benign, there would be greater confidence in the accuracy of the label compared to a single analyst label.

The model is considered low risk for evasion and adaptive attacks from sophisticated actors. In order for an adversarial attack to succeed, they would have to consistently mimic activity that generates benign alerts, tricking analysts into labelling them as such. In such cases, the model will fail due to the analyst mislabelling the alert, which was discussed as a limitation above. It is much more likely that these sophisticated attackers would be aiming to bypass the detectors that generate the alerts themselves, and is less of a concern for learning how analysts triage alerts to further prioritise them. 

The industry is now moving towards a more incident-centric model, whereby alerts that are determined to be related to a single security event are grouped into an incident before being shown to an analyst. The analyst queue then becomes a prioritised list of incidents rather than alerts. AACT could be readily expanded in that context by most simply aggregating the scores for the alerts within the incident, similarly to \cite{gelman23} where the maximum is taken over all alerts within the incident. Further research could also include holistic incident features that could highlight the severity of that incident when viewed as a whole, such as the size of the incident or the diversity of alerts and entities within the incident. This could be used to rank groups of alerts by providing a risk score for incidents, where a high score would be prioritised by analysts and a low score would be indicative that an incident is a false positive.
\section{Conclusion}\label{sec:conclusion}
In this paper, a novel approach was presented, named AACT, that automatically classifies cybersecurity alerts, by modelling short-term and long-term actions taken by SOC analysts on similar categories of alerts. It performs well at predicting the analyst triage action for alerts using features that are easy for analysts to interpret. This paper presented how this approach can be useful both for closing alerts that are not relevant and for ranking alerts that merit further inspection using a threat score. It has been successfully deployed in a production environment at \Secureworks{} and the wide applicability of the technique presented can be used by the community to enhance their security operations or for further research.

\section*{Acknowledgment}


\bibliographystyle{IEEEtran}
\bibliography{main}

\end{document}